\documentclass[]{dmtcs-episciences}

\usepackage[utf8]{inputenc}
\usepackage{subfigure}

\usepackage[round]{natbib}

\author{Walid Marweni\affiliationmark{1}\thanks{walid.marweni@gmail.com}}
\title[]{$(-k)$-critical trees and $k$-minimal trees}
\affiliation{Faculty of Sciences - Sfax University, Tunisia.}

\begin{document}

\begin{center}
    \textbf{$(-k)$-critical trees and $k$-minimal trees}\\
    Walid Marweni\\
    Faculty of Sciences - Sfax University, Tunisia.\\
    walid.marweni@gmail.com
\end{center}
~~\\
\textbf{Abstract}
In a graph $G=(V,E)$, a module is a vertex subset $M$ of $V$ such that every vertex outside $M$ is adjacent to all or none of $M$. For example, $\emptyset$, $\{x\}$ $(x\in  V )$ and $V$ are modules of $G$, called trivial modules. A graph, all the modules of which are trivial, is prime; otherwise, it is decomposable. A vertex $x$ of a prime graph $G$ is critical if $G - x$ is decomposable. Moreover, a prime graph with $k$ non-critical vertices is called $(-k)$-critical graph.
A prime graph $G$ is $k$-minimal if there is some $k$-vertex set $X$ of vertices such that there is no proper induced subgraph of $G$ containing $X$ is prime. From this perspective, I. Boudabbous proposes to find the $(-k)$-critical graphs and $k$-minimal graphs for some integer $k$ even in a particular case of graphs.
This research paper attempts to answer I. Boudabbous's question. First, it describes the $(-k)$-critical tree. As a corollary, we determine the number of nonisomorphic $(-k)$-critical tree with $n$ vertices where $k\in \{1,2,\lfloor\frac{n}{2}\rfloor\}$. Second, it provide a complete characterization of the $k$-minimal tree. As a corollary, we determine the number of nonisomorphic $k$-minimal tree with $n$ vertices where $k\leq 3$.
\section{Introduction and presentation of the results}
Our work falls within the framework of graph theory with a special focus on decomposition problems. A {\it graph} $G=(V(G), E(G))$ (or $G=(V, E)$) consists of a finite set $V$ of {\it vertices} together with a set $E$ of pairs of distinct vertices, called {\it edges}.
On the one hand, let $G=(V, E)$ be a graph. The \emph{neighborhood} of $x$ in $G$, denoted by $N_{G}(x)$ or simply $N(x)$, is the set $N_{G}(x) = \{y \in V \setminus \{x\}: \{x,y\} \in E\}$. The \emph{degree} of $x$ in $G$, denoted by $d_G(x)$ ( or $d(x)$), is the cardinal of $N_{G}(x)$.
A vertex $x$ with degree one is called a \emph{leaf}, its adjacent vertex is called a \emph{support} vertex and it is denoted by $x^+$. If $x$ is a support vertex in $G$ admitting a unique leaf neighbor, this leaf is denoted by $x^-$. The set of leaves and support vertices in a graph $G$ is denoted by $\mathcal{L}(G)$ and $\mathcal{S}(G)$ respectively.
An \emph{internal} vertex is a vertex with a degree greater than or equal to 2. The \emph{distance} between two vertices $u$ and $v$ in $G$ is the length (number of edge) of the shortest path connecting them and is denoted by $dist_G(u,v)$ or simply $dist(u,v)$. The notation $x \sim Y$ for each $Y\subseteq V\backslash \{x\}$ means $x$ is adjacent to all or none vertex of $Y$. The negation is denoted by $x \not\sim Y$.

On the other hand, given a graph $G=(V, E)$, with each subset $X$ of $V$, the graph $G[X]= (X, \displaystyle E\cap (^X_2))$ is an {\it induced subgraph} of $G$. For $X \subseteq V$ (resp. $x \in V $), the induced subgraph $G[V \setminus X]$ (resp. $G[ V\setminus \{x\}]$) is denoted by $G-X$ (resp. $G-x$). The notions of isomorphism and embedding are defined in the following way. Two graphs $G = (V, E)$ and $G' = (V', E')$ are {\it isomorphic}, which is denoted by $G \simeq G'$, if there is an {\it isomorphism} from $G$ onto $G'$, i.e., a bijection from $V$ onto $V'$ such that for all $x, y \in V$, $\{x, y\} \in E$ if and only if $\{f(x), f(y)\} \in E'$. We say that a graph $G'$ {\it embeds} into a graph $G$ if $G'$ is isomorphic to an induced subgraph of $G$. Otherwise, we say that $G$ {\it omits} $G'$.  Given a graph $G$ and a one-to-one function $f$ defined on a set containing $V(G)$, we denote by $f(G)$ the graph $(f(V(G)); f(E(G)))=(f(V(G)); \{\{f(x), f(y)\}: \{x, y\}\in E(G)\})$. A nonempty subset $C$ of $V$ is a \emph {connected component} of $G$ if for $x \in C$ and $y \in V \setminus C, \, \{x,y\} \notin E$ and if for $x \neq y \in C,$ there is a sequence $x=x_{0}, \, \ldots, \, x_{n}=y$ of $C$ elements such that for each $0\leq i \leq n-1$, $\{x_{i},x_{i+1}\} \in E$. A vertex $x$ of $G$ is \emph{isolated} if $\{x\}$ constitutes a connected component of $G$. The set of the connected components of $G$ is a partition of $V$, denoted $\mathcal{C}(G)$. The graph $G$ is \emph{connected} if it has at most one connected component of $G$. Otherwise, it is called \emph{non-connected}. For example, a \emph{tree} is a connected graph in which any two vertices are connected by exactly one path.

In addition, let's consider a graph $G=(V, E)$, a subset $M$ of $V$ is a \emph{module} of $G$ if every vertex outside $M$ is adjacent to all or none of $M$. This concept was introduced in \cite{Spinrad1992P4TreesAS} and independently under the name \emph{interval} in \cite{Cournier1998MinimalIG, Frass1984LIntervalleET, Schmerl1993CriticallyIP} and an \emph{autonomous} set in \cite{Ehrenfeucht1990PrimitivityIH}. The empty set, the singleton sets, and the full set of vertices are \emph{trivial modules}. A graph is \emph{indecomposable} (or {\it primitive}) if all its modules are trivial; otherwise, it is decomposable. Therefore, indecomposable graphs with at least four vertices are prime graphs. This concept was developed in several papers e.g (\cite{Ehrenfeucht1990PrimitivityIH, Frass1984LIntervalleET, Kelly1986InvariantsOF, Schmerl1993CriticallyIP, Sumner1973GraphsIW}), and is now elaborated in a book by Ehrenfeucht, Harju and Rozenberg \cite{Ehrenfeucht1997TheTO}. Properties of the prime substructures of a given prime structure were determined by Schmerl and Trotter \cite{Schmerl1993CriticallyIP} in their fundamental paper. Indeed, several papers within the same sphere of reference have then appeared (\cite{belkhechine2010indecomposable, belkhechine2010les, Bouchaala2013FiniteTW, DBLP:journals/arscom/Boudabbous16, Boussari2007LesG2, Ehrenfeucht1997TheTO, Ille1997IndecomposableG, Elayech2015TheD, Pouzet2009OnMP, Sayar2011PartiallyCT}). For instance, the {\it path} defined on $\mathbb{N}_n=\{1,...,n\}$, denoted by $P_n$, is prime for $n \geq 4$. A path with extremities $x$ and $y$ is referred to as $(x,y)$-path. For example, it is easy to verify that each prime graph is connected.

The study of the hereditary aspect of the primality in the graphs revolve around the following general question. Given a prime graph $G$, is there always a proper prime subgraph in $G$ ? Addressing this problematic led to the publication of numerous papers. A first result in this direction dates back to D. P. Sumner \cite{Sumner1973GraphsIW} who asserted that for every prime graph $G$, there exists $X \subseteq V (G)$ such that $|X| = 4$ and $G[X]$ is prime. In 1990, A. Ehrenfeucht and G. Rozenberg \cite{Ehrenfeucht1990PrimitivityIH} reported also that the prime graphs have the following ascendant hereditary property. Let $X$ be a subset of a prime graph $G$ such that $G[X]$ is prime. If $|V (G)\backslash X|\geq 2$, then there exist $x \neq y \in V (G) \backslash X$ such that $G[X \cup \{x, y\}]$ is prime. The latter result was improved in 1993 by J. H. Schmerl and W. T. Trotter \cite{Schmerl1993CriticallyIP} as follows: Each prime graph of order $n$, ($n \geq 7$), embeds a prime graph of order $n - 2$. It is then natural to raise the next question. Given a prime graph $G$ of order $n$, is there always a prime subgraph of $G$ of order $n-1$? The answer to this question is negative and the prime graph $G$ such that $G-x$ is decomposable for each $x \in V(G)$, referred to as \emph{critical graph}, is the counterexample.
In 1993, J.H. Schmerl and W.T. Trotter \cite{Schmerl1993CriticallyIP} characterized the critical graphs.

Consider now a prime graph $G=(V,E)$. A vertex $x$ of $G$ is said to be {\it critical} if $G-x$ is decomposable. Otherwise, $x$ is a {\it non-critical} vertex. The set of the non-critical vertices of $G$ is denoted by $\sigma(G)$. Moreover, if $G$ admits $k$ non-critical vertices, it is then called a \emph{$(-k)$-critical graph}.
Recently, Y.Boudabbous and Ille \cite{BOUDABBOUS20092839} asked about the description of the $(-1)$-critical graphs. Their question was answered by H. Belkhechine, I. Boudabbous and M. Baka Elayech \cite{belkhechine2010les} in the case of graph.
More recently, I. Boudabbous and W. Marweni described the triangle-free prime graphs having at most two non critical vertices \cite{DBLP:journals/mvl/BoudabbousM20}.

\begin{figure}[!h]
\centering
\includegraphics[width=11cm]{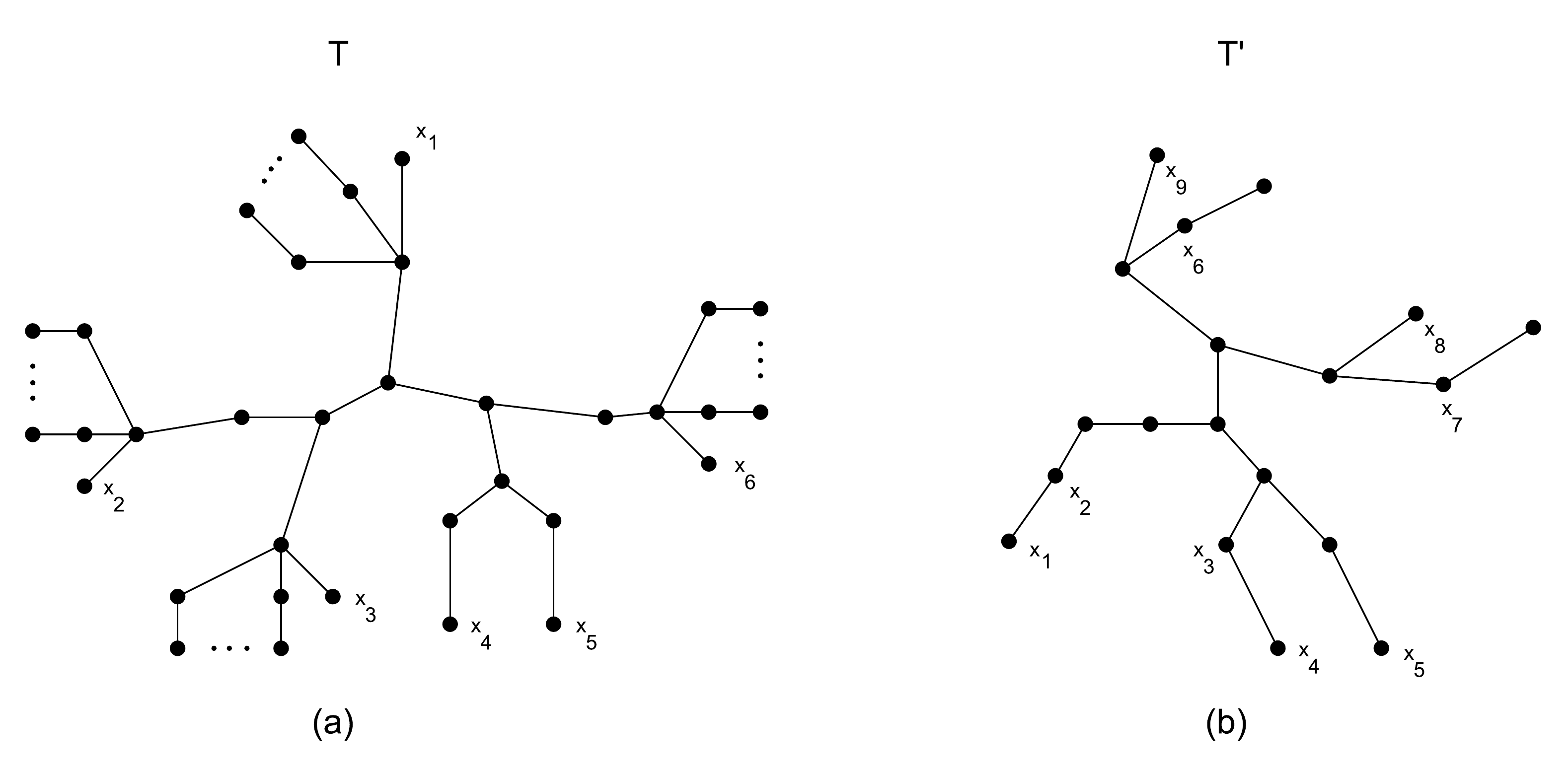}
\caption{~~$(a)$ $T$ is $(-6)$-critical and $\sigma(T)= \{x_1,...,x_6\}$.~~~~ $(b)$ $T'$ is minimal for $\{x_1,...,x_9\}$.}
\label{FK}
\end{figure}

Another intrinsic tool in this work is the notion of \emph{minimal} graphs defined as follows. A prime graph $G$ is {\it minimal} for a vertex subset $X$, or {\it $X$-minimal}, if no proper induced subgraph of $G$ containing $X$ is prime. A graph $G$ is {\it $k$-minimal} if it is minimal for some $k$-element set of $k$ elements. A. Cournier and P. Ille \cite{Cournier1998MinimalIG} in 1998 characterized the $1$-minimal and $2$-minimal graphs. Recently, M. Alzohairi and Y. Boudabbous characterized the 3-minimal triangle-free graphs \cite{ALZOHAIRI20143}. In 2015, M. Alzohairi characterized the triangle-free graphs which are minimal for some nonstable 4-vertex subset \cite{ALZOHAIRI2015159}.\\
Motivated by these two fundamental notions, I. Boudabbous proposes to find the $(-k)$-critical graphs and $k$-minimal graphs for some integer $k$ even in a particular case of graphs. This work resolves what is requested by I. Boudabbous. For this reason, we shall describe the prime tree having exactly $k$ non-critical vertices. Recall that $\lfloor x\rfloor$ denotes the greatest integer $\leq x$. Therefore, we obtain:
\begin{theorem} \label{TH10}
Let $T=(V,E)$ be a tree with at least $5$ vertices and $\{x_1,...,x_k\}$ be a vertex subset of $G$ where $k$ is an integer. \\
$T$ is $(-k)$-critical and $\sigma(T)= \{x_1,...,x_k\}$ (see Figure \ref{FK} (a)), if and only if $T$ satisfies the four assertions.
\begin{enumerate}
  \item For each $x\neq y \in \mathcal{L}(T)$, $dist(x,y)\geq 3$,
  \item $\{x_1,...,x_k\} \subseteq \mathcal{L}(T)$ and $1 \leq k\leq \lfloor\frac{n}{2}\rfloor$,
  \item For each $x \in \mathcal{L}(T)\backslash\{x_1,...,x_k\}$, there is a unique $i\in\{1,...,k\}$ such that $dist(x,x_i)=3$ and $d(x^+)=2$,
  \item If $d(x_{i}^+)=2$ where $i \in \{1,...,k\}$, then for all $x \in \mathcal{L}(T)\backslash\{x_i\}$, $dist(x_i, x)\geq 4$.
\end{enumerate}
\end{theorem}
Moreover, we shall describe the $k$-minimal trees. As a matter of fact, we obtain:
\begin{theorem} \label{TH20}
Let $T=(V,E)$ be a tree with at least $5$ vertices and let $\{x_1,...,x_k\}$ be a vertex subset of $G$ where $k$ is a strictly positive integer.\\
$T$ is minimal for $\{x_1,...,x_k\}$ (see Figure \ref{FK} (b)) if and only if $T$ satisfies the three assertions.
\begin{enumerate}
  \item For each $x\neq y \in \mathcal{L}(T)$, $dist(x,y)\geq 3$.
  \item For each $x \in \mathcal{L}(T)$, $\{x, x^+\}\cap\{x_1,...,x_k\}\neq\emptyset$.
  \item If $x_i\in \mathcal{S}(T)$ and $x_i^-\notin \{x_1,...,x_k\}$ where $i \in \{1,...,k\}$, then $d(x_i)=2$ and there is $j\neq i \in \{1,...,k\}$ such that $x_j \in \mathcal{L}(T)$ and $d(x_i, x_j)=2$.
\end{enumerate}
\end{theorem}

\section{Proof of Theorem 1.1:}

We recall the characterization of the prime tree set forward by to M. Alzohairi and Y. Boudabbous.
\begin{lemma}{\rm (\cite{ALZOHAIRI20143})}\label{tree}
\begin{enumerate}
  \item If $M$ is a nontrivial module in a decomposable tree $T$, then $M$ is a stable set of $T$. Moreover, the elements of $M$ are leaves of $T$.
  \item A tree with at least four vertices is prime if and only if any two distinct leaves do not have the same neighbor.
\end{enumerate}
\end{lemma}
As an immediate consequence of Lemma \ref{tree}, we have the following result.
\begin{corollary}\label{wtree}
Let $T=(V,E)$ be a tree. $T$ is prime if and only if $d(x,y)\geq 3$, for each $x\neq y \in \mathcal{L}(T)$.
\end{corollary}
The following observation follows immediately from Lemma \ref{tree}.
\begin{observation}\label{O1}
Let $T=(V,E)$ be a prime tree with $n$ vertices. Then, $|\mathcal{S}(T)|=|\mathcal{L}(T)|\leq \lfloor\frac{n}{2}\rfloor$.
\end{observation}
Now, we establish the next lemma that will be needed in the sequel.
\begin{lemma} \label{LL1}
Let $T=(V,E)$ be a prime tree and $x\in \mathcal{L}(T)$. If $T-x$ is decomposable, then there is $y \in \mathcal{L}(T)\backslash\{x\}$ such that $\{y, x^+\}$ is the unique module of $T-x$.
\end{lemma}
\noindent{\textbf{Proof:}} Consider a prime tree $T=(V,E)$ and $x\in \mathcal{L}(T)$. Assume that $T-x$ is a decomposable tree. Resting upon Lemma \ref{tree}, there exist two distinct leaves of $T-x$, said $y$ and $z$, which have the same neighbor. Hence, $\{y, z\}$ is a module of $T-x$. Since $T$ is a prime tree, $x \not\sim \{y,z\}$. Thus, $x^+\in\{y,z\}$. Without loss of generality, we may assume that $x^+ =z$ and we have $y\in \mathcal{L}(T)$. Since $T$ is prime and $y\in \mathcal{L}(T)$, then $dist(y,u)\geq 3$ for each $u \neq y\in \mathcal{L}(T)$. Therefore, $\{y, x^+\}$ is the unique module of $T-x$.{\hspace*{\fill}$\Box$\medskip}\\
\noindent{\textbf{Proof of Theorem \ref{TH10}.}} Consider a tree $T=(V,E)$ with $n$ vertices where $n\geq 5$ and $\{x_1,...,x_k\}$ is a subset of $V$ where $k$ is a strictly positive integer.\\
Assume that $T$ is $(-k)$-critical and $\sigma(T)= \{x_1,...,x_k\}$. Since $T$ is prime, by Corollary \ref{wtree}, we have for each $x \neq y \in \mathcal{L}(T),~dist(x,y)\geq 3$. Hence, $T$ satisfies the condition (1) of Theorem \ref{TH10}.\\
Moreover, let $x\in V \backslash \mathcal{L}(T)$, $x$ is an internal vertex of $T$ and $T-x$ is a non-connected graph. Then, $T-x$ is decomposable and $x\notin \sigma(T)$. Thus, $\{x_1,...,x_k\} \subseteq \mathcal{L}(T)$. As $T$ is prime, based on Observation \ref{O1}, we have $1\leq k\leq \lfloor\frac{n}{2}\rfloor$. Hence, $T$ satisfies condition (2) of Theorem \ref{TH10}.\\
Now, consider $x \in \mathcal{L}(T)\backslash\{x_1,...,x_k\}$. Then, $T-x$ is a decomposable tree. By Lemma \ref{LL1}, there is $y \in \mathcal{L}(T)\backslash\{x\}$ such that $\{y, x^+\}$ is the only module of $T-x$. Clearly,  $dist(x,y)=3$ and $d(x^+)=2$. Now, prove that $y\in \sigma(T)$. To the contrary, suppose that $y\notin \sigma(T)$, implying that $T-y$ is a decomposable tree. Using again Lemma \ref{LL1},
there is $z \in \mathcal{L}(T)\backslash\{y\}$ such that $\{z, y^+\}$ is the unique module of $T-y$. Thus, $d(y^+)=2$ and $dist(y,z)=3$. This implies that $z=x$ and we obtain that $T$ is with 4 vertices; which contradicts the fact that $T$ is a tree having at least 5  vertices. Hence, $y\in \sigma(T)$. Therefore, $T$ satisfies the condition (3) of Theorem \ref{TH10}.\\
Besides, assume that there is $i \in \{1,...,k\}$ such that $d(x_i^+)=2$. Then, $T-x_i$ is a prime tree and $x_i^+ \in \mathcal{L}(T-x_i)$. Referring to Corollary \ref{wtree}, for all $y\in\mathcal{L}(T-x_i)$, $dist(x_i^{+}, y)\geq3$. Since $\mathcal{L}(T-x_i)\backslash \{x_i^+\}=\mathcal{L}(T) \backslash \{x_i\}$, then for each $y\neq x_i\in \mathcal{L}(T)$, $dist(x_i,y)\geq 4$. Hence, $T$ satisfies condition (4) of Theorem \ref{TH10}.

Conversely, assume that $T$ satisfies the conditions (1)-(4) of Theorem \ref{TH10}. Proving that, $T$ is $(-k)$-critical and $\sigma(T)= \{x_1,...,x_k\}$. Since for each $x\neq y \in \mathcal{L}(T)$, $dist(x,y)\geq 3$ and by Corollary \ref{wtree}, $T$ is prime. Clearly, if $x\in V \backslash \mathcal{L}(T)$, $T-x$ is a non-connected graph. Thus, $T-x$ is decomposable and hence $x$ is a critical vertex.\\
Furthermore, if $x \in \mathcal{L}(T)\backslash\{x_1,...,x_k\}$, by assertion (3), there is a unique $i\in\{1,...,k\}$ such that $dist(x,x_i)=3$ and $d(x^+)=2$. Then, $\{x^{+}, x_i\}$ is a module of $T-x$. Hence for each $x\in\mathcal{L}(T)\backslash\{x_1,...,x_k\}$, $T-x$ is decomposable and then $x$ is a critical vertex.\\
Now, given $i \in \{1,...,k\}$; if $d(x^+_i)\geq 3$, then $\mathcal{L}(T)\backslash\{x_i\}=\mathcal{L}(T-x_{i})$. According to first hypothesis of Theorem \ref{TH10}, for each $x\neq y \in\mathcal{L}(T-x_{i})$, $dist(x, y)\geq3$. By Corollary \ref{wtree}, $T-x_i$ is a prime tree.\\
Assume now that $d(x_i^+)=2$. Suppose that $T-x_i$ is a decomposable tree. Then, by Lemma \ref{LL1} there is a unique $y \neq x_i \in \mathcal{L}(T)$ such that $\{y, x_i^+\}$ is the unique module of $T-x_i$. As a matter of fact, $dist(y,x_i)=3$; which contradicts the hypothesis 4 of Theorem \ref{TH10}. Hence, $T-x_i$ is prime. Consequently, $T$ is $(-k)$-critical and $\sigma(T)= \{x_1,...,x_k\}$.
{\hspace*{\fill}$\Box$\medskip}

\begin{figure}[h!]
\centering
\includegraphics[width=5cm]{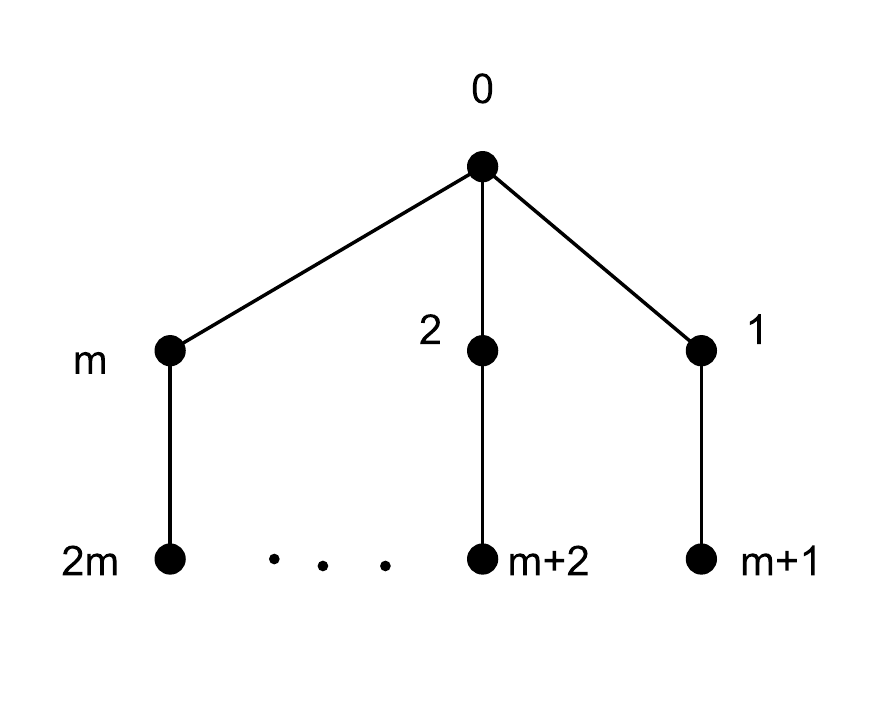}
\caption{The tree $A_{2m+1}$}
\label{AA}
\end{figure}

\begin{figure}[h!]
\centering
\includegraphics[width=12cm]{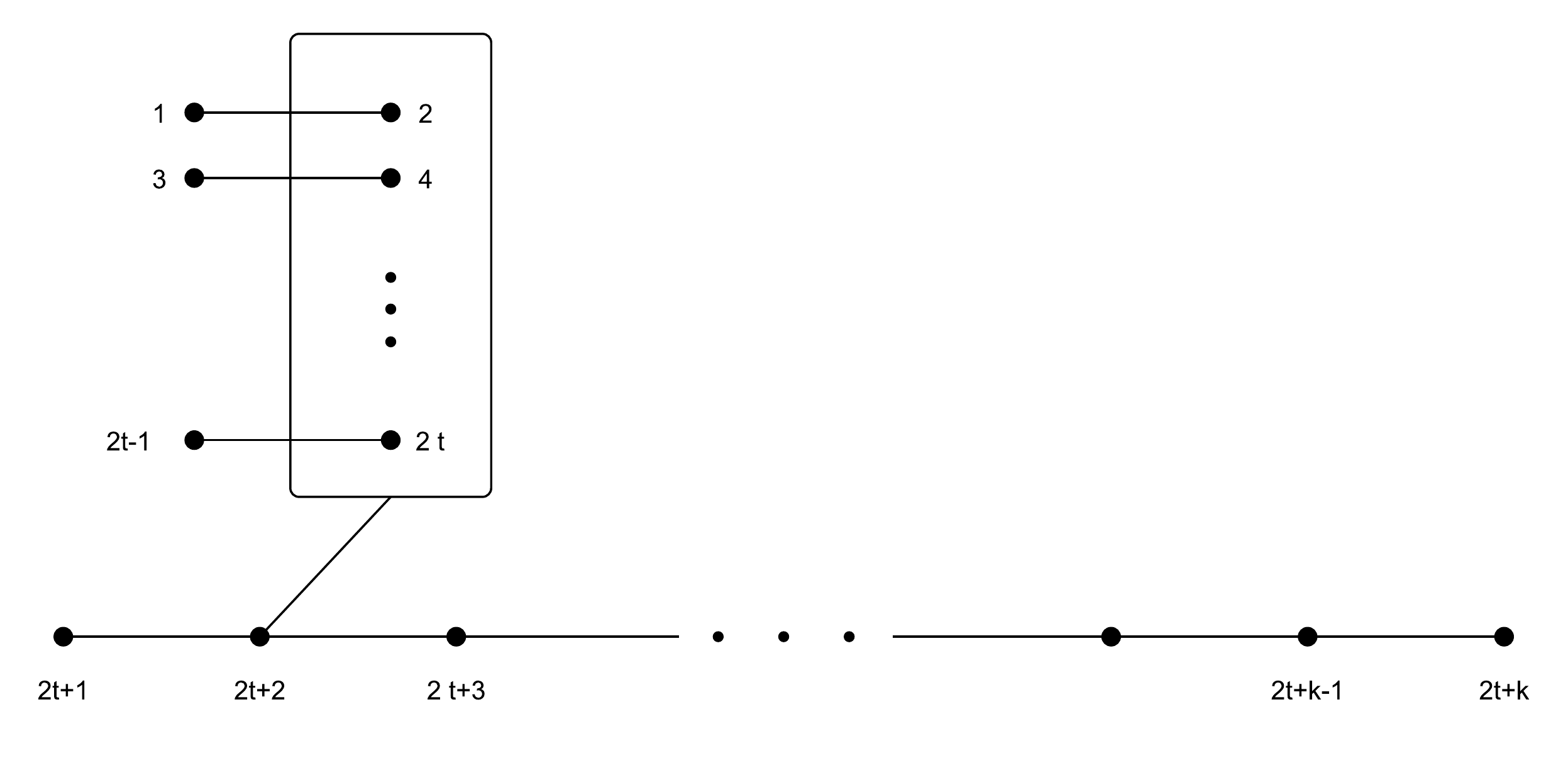}
\caption{The tree $P_{k, t}$}
\label{2CR1}
\end{figure}

\begin{figure}[h!]
\centering
\includegraphics[width=12cm]{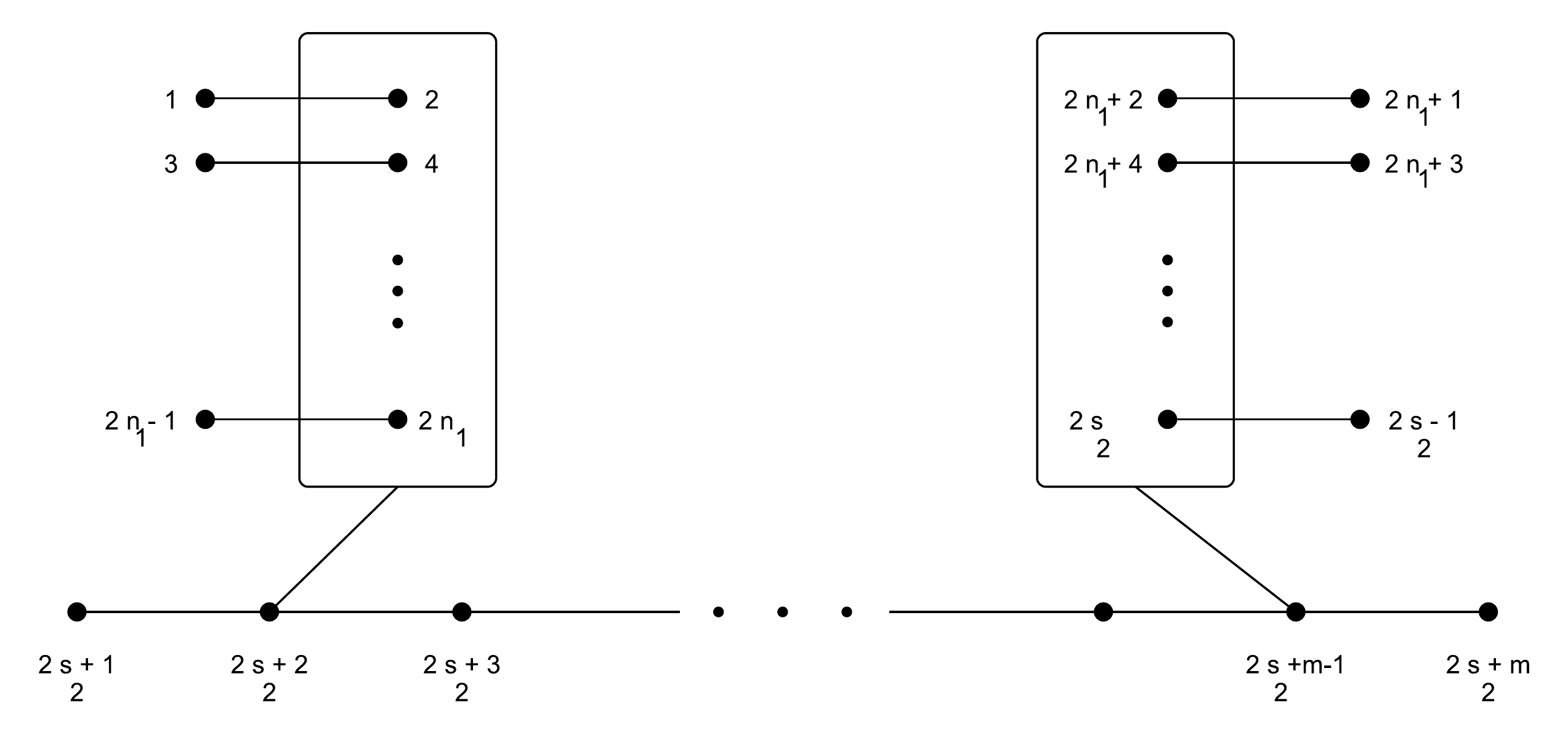}
\caption{The tree $P_{m, n_1,n_2}$}
\label{2CR}
\end{figure}

Our second objective in this section lies in determining the number of nonisomorphic $(-k)$-critical trees with $n\geq5$ vertices where $k \in \{1,2, \lfloor\frac{n}{2}\rfloor\}$. According to the characterization of critical graphs \cite{Schmerl1993CriticallyIP}, $P_4$ is the a unique critical trees. To specify the the number of nonisomorphic $(-k)$-critical trees where $k \in \{1,2,\lfloor\frac{n}{2}\rfloor\}$, we introduce for all $n\in \mathbb{N}$, the one-to-one function:
$$\begin{array}{ccccc}
T_n & : & \mathbb{N} & \to & \mathbb{N} \\
 & & p & \mapsto & p+n \\
\end{array}$$
Now, we introduce also the following trees.
\vspace{-0.2cm}
\begin{description}
\item[$\bullet$] For integers $m\geq 2$, let $A_{2m+1}$ be the tree defined on $\{0,...,2m\}$ and $E(A_{2m+1})=\{\{0,i\}, \{i,i+m\} : 1\leq i \leq m\}$ (see Figure \ref{AA}).
  \item[$\bullet$] For integers $k\geq 4$, $t\geq 1$, let $P_{k, t}$ be the tree defined on $\{1,...,2t+k\}$ and $E(P_{k,t})=E(T_{2t}(P_k))\cup\{\{2i-1, 2i\}: 1\leq i \leq t\}\cup \{\{2t+2, 2i\}: 1\leq i \leq t\}$ (see Figure \ref{2CR1}).
  \item[$\bullet$] For integers $m\geq 4$, $n_1\geq 1$, $n_2\geq 1$, for each $p\in \{1,2\}$, $s_p=n_1+...+n_p$. Let $P_{m, n_1,n_2}$ be the tree defined on $\{1,...,2s_2+m\}$ and $E(P_{m,n_1,n_2})=E(T_{2s_2}(P_m))\cup\{\{2i-1, 2i\}: 1\leq i \leq s_2\}\cup \{\{2s_2+2, 2i\}, \{2s_2+m-1, 2j\}: 1\leq i \leq n_1 ~~and ~~n_1< j \leq s_2\}$ (see Figure \ref{2CR}).
\end{description}
\begin{proposition}
\begin{enumerate}
  \item Up to isomorphisms, the $(-1)$-critical trees with $n$ vertices are the tree $\displaystyle P_{4,\frac{n-4}{2}}$ where $n$ is an even integer $\geq 6$.
  \item Up to isomorphisms, the $\displaystyle (-\lfloor\frac{n}{2}\rfloor)$-critical trees with $n$ vertices are the tree $A_{n}$ where $n$ is an odd integer $\geq 5$.
  \end{enumerate}
\end{proposition}
\noindent{\textbf{Proof:}}\begin{enumerate}
  \item Clearly, departing from Theorem \ref{TH10}, $\displaystyle P_{4,\frac{n-4}{2}}$ is a $(-1)$-critical tree where $n\geq 6$ and $\displaystyle\sigma(P_{4,\frac{n-4}{2}})=\{n-3\}$. Now, we consider a $(-1)$-critical tree $T$ with $n\geq 5$ vertices such that $\sigma(T)=\{x_1\}$. By Theorem \ref{TH10}, $x_1 \in \mathcal{L}(T)$. If $x\neq x_1\in \mathcal{L}(T)$, then by assertion (3) of Theorem \ref{TH10}, $dist(x, x_1)=3$ and $d(x^+)=2$. On the contrary, suppose that $|\mathcal{L}(T)|=2$. Since $T$ is prime, then $T$ is isomorphic to $P_4$; which contradicts the fact that $T$ has at least 5 vertices. Hence, $|\mathcal{L}(T)|\geq 3$. By assertion (3), for each $y\in\mathcal{L}(T)$, $dist(y, x_1)=3$ and $d(y^+)=2$. Thus, $T$ is isomorphic to $\displaystyle P_{4,\frac{n-4}{2}}$ where $n\geq 6$ is an even integer.
  \item By Theorem \ref{TH10}, $A_{2m+1}$ is $\displaystyle(-\lfloor\frac{2m+1}{2}\rfloor)$-critical and $\sigma(A_{2m+1})=\mathcal{L}(A_{2m+1})$ where $m\geq 2$. Now, we consider a $\displaystyle(-\lfloor\frac{n}{2}\rfloor)$-critical tree $T$ with $n\geq 5$ vertices. Using Theorem \ref{TH10}, $\sigma(T)\subseteq \mathcal{L}(T)$ implies that $\displaystyle|\sigma(T)|=\lfloor\frac{n}{2}\rfloor \leq |\mathcal{L}(T)|$. By Observation \ref{O1}, $\displaystyle|\mathcal{L}(T)| =|\mathcal{S}(T)|\leq \lfloor\frac{n}{2}\rfloor$ and therefore $\displaystyle|\mathcal{L}(T)| =|\mathcal{S}(T)|=\lfloor\frac{n}{2}\rfloor$. Now, we shall prove that $n$ is odd. To the contrary, suppose that $n$ is even. Then, $\displaystyle|\mathcal{L}(T)| =|\mathcal{S}(T)|=\frac{n}{2}$. Hence, $V(T)=\mathcal{L}(T)\cup\mathcal{S}(T)$. Since $T[\mathcal{S}(T)]$ is a tree, there exists a vertex $y\in \mathcal{S}(T)$ with $d_{T[\mathcal{S}(T)]}(y)=1$. Hence, $d_{T}(y)=2$. We may assume that $N(y)=\{y^-,x\}$ where $x\in \mathcal{S}(T)$. Thus, $\{y, x^-\}$ is a module of $T-y^-$; which contradicts the fact that $y^-$ is not a critical vertex. Accordingly, $n$ is odd, $\displaystyle |\mathcal{L}(T)|=|\mathcal{S}(T)|=\frac{n-1}{2}$, and $V(T)=\mathcal{L}(T)\cup\mathcal{S}(T)\cup\{z\}$. We can assume that $\displaystyle\mathcal{L}(T)=\{x_1,...,x_{(\frac{n-1}{2})}\}$ and $\displaystyle\mathcal{S}(T)=\{x^+_1,...,x^+_{(\frac{n-1}{2})}\}$.
  Since $T[\mathcal{S}(T)\cup\{z\}]$ is a tree, there exists a vertex $x_i^+\in \mathcal{S}(T)$ where $\displaystyle 1\leq i \leq \frac{n-1}{2}$ with $d_{T[\mathcal{S}(T)]}(x_i^+)=1$ and hence $d_{T}(x_i^+)=2$. By Theorem \ref{TH10}, $dist(x_i, x_j)\geq 4$ for each $\displaystyle j\neq i \in \{1,...,\frac{n-1}{2}\}$. Hence, $T$ is isomorphic to $A_{n}$ where $n\geq 5$.{\hspace*{\fill}$\Box$\medskip}
\end{enumerate}

As a consequence of Theorem \ref{TH10}, we get the following result.
\begin{proposition}\label{p1}
Up to isomorphisms, the $(-2)$-critical trees with $n\geq 5$ vertices are the trees $P_n$, $P_{k,t}$ where $k\geq 4$, $t\geq 1$ and $n=k+2t$, and $P_{m, n_1, n_2}$ where $m\geq4$, $n_1, n_2\geq 1$ and $n=m+2(n_1+n_2)$.
\end{proposition}\label{2CT}
\noindent{\textbf{Proof:}}
By Theorem \ref{TH10}, $P_n$, $P_{k,t}$ where $k\geq 4$, $t\geq 1$, and $P_{m, n_1, n_2}$ where $m\geq4$, $n_1, n_2\geq 1$ are $(-2)$-critical trees and $\sigma(P_n)=\{1, n\}$, $\sigma(P_{k,t})=\{2t+1,2t+k\}$, and $\sigma(P_{m, n_1, n_2})=\{2s_2 +1, 2s_2+ m\}$. Now, assume that $T$ is a $(-2)$-critical tree with $n\geq 5$ vertices such that $\sigma (T)=\{x_1,x_2\}$. By Theorem \ref{TH10}, $x_1, x_2 \in \mathcal{L}(T)$. As $T$ is a prime tree, then the $(x_1,x_2)$-path is isomorphic to $P_k$ where $k \geq4$.
If $|\mathcal{L}(T)|=2$, then $T$ is isomorphic to $P_n$ and $n=k$. Assume that $|\mathcal{L}(T)|\geq 3$, then by Theorem \ref{TH10}, for each $x\in \mathcal{L}(T)\backslash\{x_1, x_2\}$; there is a unique $i\in \{1,2\}$ such that $dist(x, x_i)=3$ and $d(x^+)=2$. Hence, $T$ is isomorphic to $P_{k,t}$ where $k\geq 4$, $t\geq 1$ and $n=k+2t$ or $T$ is isomorphic to $P_{k, n_1, n_2}$ where $k\geq4$, $n_1, n_2\geq 1$ and $n=k+2(n_1+n_2)$.{\hspace*{\fill}$\Box$\medskip}

\begin{theorem} \label{TH2}
The number of nonisomorphic $(-2)$-critical trees with $n$ vertices equals:\\
$\bullet$ $\displaystyle\left\lfloor\frac{n}{4}\right\rfloor^{2}-1$ if $n\equiv 0~(mod~4)$.\\
$\bullet$ $\displaystyle\left\lfloor\frac{n}{4}\right\rfloor^{2}$ if $n\equiv 1~(mod~4)$.\\
$\bullet$ $\displaystyle\left\lfloor\frac{n}{4}\right\rfloor\left(\left\lfloor\frac{n}{4}\right\rfloor+1\right)-1$ if $n\equiv 2~(mod~4)$.\\
$\bullet$ $\displaystyle\left\lfloor\frac{n}{4}\right\rfloor\left(\left\lfloor\frac{n}{4}\right\rfloor+1\right)$ otherwise.
\end{theorem}
\noindent{\textbf{Proof:}} At the beginning, it is not difficult to verify that there are not two isomorphic different trees in the
union $\{P_m: m\geq 5\}\cup\{P_{k,t}: k\geq5, t\geq 1\}\cup\{P_{m,n_1,n_2}: m\geq4, n_1\geq1 ~~and~~ n_2\geq 1\}$. \\
By Proposition \ref{p1}, $P_5$ is the unique $(-2)$-critical tree with five vertex and $P_6$ is the unique $(-2)$-critical tree with six vertices. As a matter of fact, the result holds.\\
Now, assume that $n\geq 7$. By Proposition \ref{2CT}, the nonisomorphic $(-2)$-critical trees with $n$ vertices are $P_n$, the family of $P_{k,t}$ where $t \geq1$, $k\geq5$, and $n=2t+k$, or the family of $P_{m, n_1, n_2}$ where $1\leq n_1 \leq n_2$, $m\geq4$, and $n=2(n_1+n_2)+m$.
Therefore, it is sufficient to determine the number of the family of $P_{k,t}$ and the number of the family of $P_{m, n_1, n_2}$.\\
Let $S_{m}=\{(n_1, n_2)\in \mathbb{N}\times \mathbb{N}: 1\leq n_1\leq n_2, ~~n_1+n_2=\frac{n-m}{2}\}$, where $4\leq m\leq n-4$ and let $C_{t}=\{k\in \mathbb{N}: 5\leq k~~and ~~k=n-2t\}$, where $1\leq t\leq \frac{n-5}{2}$.
Since $n-m= 2(n_1+n_2)$, it is obvious that $n$ and $m$ are of the same parity. Hence, we distinguish two cases.\\
$\underline{\textbf{Case 1:}}$ If $n=2p$ where $4 \leq p$ and $m=2q$ where $2 \leq q \leq p-2$. \\
Consider $\displaystyle S=\bigcup_{q=2}^{p-2}S_{2q}$ and $\displaystyle C=\bigcup_{t=1}^{p-3}C_t$. First, it is clear that the number of the family of $P_{k,t}$ is the cardinality of the set $C$. Moreover, it is clear that $|C_t|=1$ where $1\leq t\leq p-3$. Hence, $\displaystyle |C|=\sum_{t=1}^{p-3}|C_t|=p-3$. Second, obviously the number of the family of $P_{m, n_1, n_2}$ is the cardinality of the set $S$. Furthermore, we have $\displaystyle |S|=\sum_{q=2}^{p-2}|S_{2q}|$. It is noticeable that for each $2\leq q \leq p-2$,  $|S_{2q}|= P_2(\frac{n-2q}{2})$, where $P_i(j)$ is the number of partitions of $j$ to $i$ parts. Recall that for an integer $k \geq 3$, $P_2(k) =\lfloor\frac{k}{2}\rfloor$ where $\lfloor x \rfloor$ is the greatest integer $\leq x$ \cite{anderson_2006}.
We get then $$\begin{aligned}
\displaystyle |S| &= \sum_{q=2}^{p-2} P_{2}\left(\frac{n-2q}{2}\right)~~\\
                  & = \sum_{q=2}^{p-2} \left\lfloor\frac{n-2q}{4}\right\rfloor \\
                  &  =\sum_{q=2}^{p-2} \left\lfloor\frac{\frac{n}{2}-q}{2}\right\rfloor\\
                  & =\sum_{i=0}^{p-2} \left\lfloor\frac{i}{2}\right\rfloor.\\
                  & =\displaystyle \left\{
   \begin{array}{ll}
     ~\displaystyle (k-1)^2 & if ~p=2k, \hbox{} \\
     ~~\\
  ~\displaystyle (k-1)k & if ~p=2k+1. \hbox{}
   \end{array}
 \right.
\end{aligned}$$

$\underline{\textbf{Case 2:}}$ Let $n=2p+1$ where $4 \leq p$ and $m=2q+1$ where $2 \leq q \leq p-2$.
Let $\displaystyle S=\bigcup_{q=2}^{p-2}S_{2q+1}$ and $\displaystyle C=\bigcup_{t=1}^{p-2}C_t$. Clearly, the number of the family of $P_{k,t}$ is the cardinality of the set $C$. Hence, $\displaystyle |C|=\sum_{t=1}^{p-2}|C_t|=p-2$.
In addition, the number of the family of $P_{m, n_1, n_2}$ is the cardinality of the set $S$.
Therefore, we have $\displaystyle |S|=\sum_{q=2}^{p-2}|S_{2q+1}|$.\\
Since for each $2\leq q \leq p-2$,  $|S_{2q+1}|= P_2\left(\frac{(n-1)-2q}{2}\right)$. Proceeding in the same manner as case 1, if $p=2k$ where $k\geq 2$, then $|S|=(k-1)^2$. Otherwise, $|S|=(k-1)k$.\\
Consequently, the number of nonisomorphic $(-2)$-critical trees with $n$ vertices equals:
$$\displaystyle \left\{
   \begin{array}{ll}
~\displaystyle \left\lfloor\frac{n}{4}\right\rfloor^{2}-1 & if ~n\equiv 0~(mod~4), \hbox{} \\
     ~~\\
~\displaystyle\left\lfloor\frac{n}{4}\right\rfloor^{2}  &if ~n\equiv 1~(mod~4),  \hbox{} \\
~~\\
~\displaystyle \left\lfloor\frac{n}{4}\right\rfloor\left(\left\lfloor\frac{n}{4}\right\rfloor+1\right)-1& if ~n\equiv 2~(mod~4),  \hbox{}\\
~~\\
~\displaystyle\left\lfloor\frac{n}{4}\right\rfloor\left(\left\lfloor\frac{n}{4}\right\rfloor+1\right) &if ~n\equiv 3~(mod~4).  \hbox{}
   \end{array}
 \right.
$$
{\hspace*{\fill}$\Box$\medskip}

\section{Proof of Theorem 1.2:}
We set two major objectives throughout this section. First, to characterize the $k$-minimal trees. Second, to determine the number of nonisomorphic $k$-minimal trees with $n$ vertices where $k \in \{1,2,3\}$.

\noindent{\textbf{Proof of Theorem 1.2.} Let $T=(V,E)$ be a tree with $n\geq 5$ vertices and let $\{x_1,...,x_k\}$ be a vertex subset of $G$ where $k$ is a strictly positive integer. Assume that $T$ is minimal for $\{x_1,...,x_k\}$. Since $T$ is prime, it satisfies the first condition of Theorem \ref{TH20}. Suppose, on the contrary, that there is $y \in \mathcal{L}(T)$ such that $\{y, y^+\}\cap \{x_1,...,x_k\}=\emptyset$. Since $T-y$ is a decomposable tree, by Lemma \ref{LL1}, there is $x\in \mathcal{L}(T)$ such that $\{x, y^+\}$ is a module of $T-y$. Thus, $d(y^+)=2$ and $d(x, y)=3$. If $x\notin \{x_1,...,x_k\}$, then $T-x$ is a decomposable tree. By using again Lemma \ref{LL1}, $d(x^+)=2$ and so $T$ is isomorphic to $P_4$; which contradicts the fact that $n\geq 5$.\\
Moreover, assume that $x\in \{x_1,...,x_k\}$ and $d(x^+) \geq 3$, then $\mathcal{L}(T)\backslash\{y\}=\mathcal{L}(T-\{y, y^+\})$. By Lemma \ref{tree}, $T-\{y, y^+\}$ is a prime tree containing $\{x_1,...,x_k\}$; which contradicts the fact that $T$ is minimal for $\{x_1,...,x_k\}$. Hence, for each $x\in \mathcal{L}(T)$, $\{x, x^+\}\cap\{x_1,...,x_k\}\neq\emptyset$ and $T$ satisfies the second condition of Theorem \ref{TH20}.\\
Now, assume that there is $x_i\in \mathcal{S}(T)$ and $x_i^-\notin \{x_1,...,x_k\}$ where $1\leq i \leq k$. On the contrary, suppose that $d(x_i) \geq 3$, then $\mathcal{L}(T)\backslash\{x_i^-\}=\mathcal{L}(T-x_i^-)$. By Lemma \ref{tree}, $T-x_i^-$ is a prime tree containing $\{x_1,...,x_k\}$; which is impossible. Hence, $d(x_i)=2$. On the contrary, suppose that for each $j\neq i \in \{1,...,k\}$ such that $x_j \in \mathcal{L}(T)$, $d(x_i, x_j)\geq 3$. Since $T-x_i^-$ is a decomposable tree, then by Lemma \ref{LL1} there is $y\in \mathcal{L}(T)$ such that $d(x_i, y) =2$ and hence $y\notin \{x_1,...,x_k\}$. By Condition 2, $y^+ \in \{x_1,...,x_k\}$ and so $d(y^+)=2$. Thus, $T$ is isomorphic to $P_4$; which is impossible. Therefore, $T$ satisfies the third condition.

Conversely, let $T=(V,E)$ be a tree with $n\geq 5$ vertices. Since for each $x\neq y \in \mathcal{L}(T)$, $d(x, y)\geq 3$, $T$ is a prime tree. Let $X$ be a subset of $V$ such that $\{x_1,...,x_k\} \subseteq X$ and $T[X]$ is prime. Consider $x\in \mathcal{L}(T)$. If $x\in \{x_1,...,x_k\}$, then $x\in X$.\\
Now, assume that $x\notin \{x_1,...,x_k\}$. On the contrary, suppose that $x\notin X$. By assertion (2) of Theorem \ref{TH20}, $x^+\in \{x_1,...,x_k\}$. Since $T$ satisfies assertion (3) of Theorem \ref{TH20}, then $d(x^+)=2$ and there is $i \in \{1,...,k\}$ such that $x_i\in \mathcal{L}(T)$ and $d(x_i, x^+)=2$. Thus, $\{x^+, x_i\}$ is a module of $T[X]$; which is impossible. Hence, $x\in X$. We conclude that $\mathcal{L}(T)\subset X$. Since $T[X]$ is a prime, it is connected.  Therefore, $T[X]$ is a tree containing $\mathcal{L}(T)$. Hence, $X=V$. Thus, $T$ is minimal for $\{x_1,...,x_k\}$.{\hspace*{\fill}$\Box$\medskip}

The following corollary is an immediate consequence of Theorem \ref{TH20}.
\begin{corollary}
For any distinct vertices $x_1, x_2$,..., and $x_k$ in a prime tree $H$, there is an induced subtree $T$ of $H$ that contains $\{x_1, x_2,...,x_k\}$, and satisfies the assertions of Theorem \ref{TH20}.
\end{corollary}

Our second objective is to determine the number of nonisomorphic $k$-minimal trees with $n$ vertices where $k \in \{1,2,3\}$. According to the characterization of 1-minimal and 2-minimal graphs, $P_4$ is the unique 1-minimal tree and $P_k$, where $k\geq 4$, is the unique 2-minimal tree \cite{Cournier1998MinimalIG}.

To specify the number of nonisomorphic $3$-minimal trees with $n$ vertices, we introduce the following tree.
\begin{description}
  \item[$\bullet$] For positive integers $k$, $m$, $n$ with $k \leq m \leq n$, let $S_{k,m,n}$ be the $(k+m+n+1)$-vertex tree with the union of the paths of lengths $k$, $m$, and $n$ having common endpoint $r$. Let $a_1,...,a_k$, $b_1,...,b_m$, and $c_1,...,c_n$ denote the other vertices on these paths, indexed by their distance from $r$ (see Figure \ref{3MINM}).
\end{description}
\begin{figure}[h!]
\centering
\includegraphics[width=10cm]{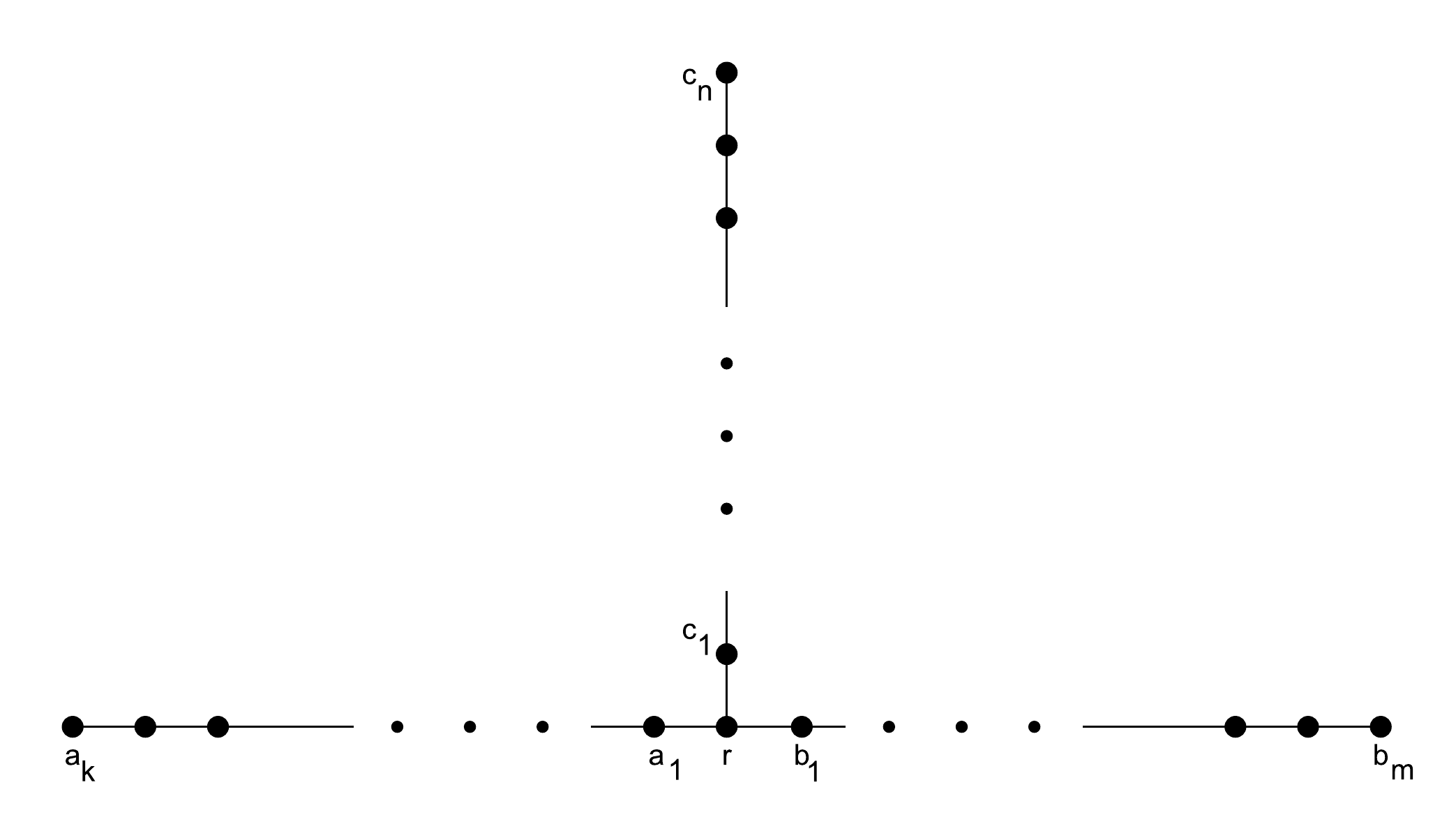}
\caption{$S_{k,m,n}$}
\label{3MINM}
\end{figure}

As an immediate consequence of Theorem \ref{TH20}, we get the following result which was already obtained by M. Alzohairi and Y. Boudabbous in \cite{ALZOHAIRI20143}.
\begin{corollary}\label{3mw}
Let $x$, $y$, and $z$ be distinct vertices in a tree $T$. The tree $T$ is minimal for $\{x,y,z\}$ if and only if $T$ and $\{x,y,z\}$ have one of the following forms:
\begin{enumerate}
\item $T \simeq P_4$.
\item $T \simeq P_k$ with $k\geq5$ such that $\{x,y,z\}$ contains the leaves.
\item $T \simeq S_{k,m,n}$ with $m\geq 2$ such that $x$, $y$, and $z$ are the leaves.
\item $T \simeq S_{1,2,n}$ such that $\{x,y,z\}=\{a_1, b_1, c_n\}$.
\item $T \simeq S_{1,2,2}$ such that $\{x,y,z\} = \{a_1, b_1, c_1\}$.
\end{enumerate}
\end{corollary}
\begin{proposition}
The number of nonisomorphic $3$-minimal trees with $n$ vertices equals:\\
$\bullet$ 1 if $n\in \{4,5\}$.\\
$\bullet$ 2 if $n=6$.\\
$\bullet$ $\left[\frac{(n-1)^2}{12}\right]-\left\lfloor\frac{n-4}{2}\right\rfloor + \left\lfloor\frac{n-2}{2}\right\rfloor-1$ if $n\geq 7$, where $[x]$ is the nearest integer from $x$.
\end{proposition}
\noindent{\textbf{Proof:}} At the beginning, it is not difficult to verify that there are not two isomorphic different graphs in the union $\{P_k: k\geq 4\}\cup \{S_{k,m,n} : m=2\}$. \\
By Corollary \ref{3mw}, $P_4$ is the unique 3-minimal tree with four vertices and $P_5$ is the unique 3-minimal tree with five vertices. In addition, the only $3$-minimal tree with six vertices are $P_6$ and $S_{1,2,2}$. Therefore, the result holds for $n \in \{4,5,6\}$.\\
Now, assume that $n\geq 7$. By Corollary \ref{3mw}, the non isomorphic $3$-minimal $n$-vertex trees are $P_n$ and the family of $S_{k,m,t}$, where $k \leq m\leq t$, $m\geq 2$, and $k+m+t+1=n$. Therefore, it is sufficient to prove that the cardinality of the set $S=\{(k,m,t)\in \mathbb{N}\times\mathbb{N}\times\mathbb{N}:1\leq k \leq m\leq t, m\geq 2, k+m+t=n-1\}$ equals $$\displaystyle \left[\frac{(n-1)^2}{12}\right]-\left\lfloor\frac{n-4}{2}\right\rfloor+\left\lfloor\frac{n-2}{2}\right\rfloor-2.$$
Let $S_2=\{(k,m,t)\in S: k\geq 2\}$. It is easy to infer that $|S-S_2|=P_2(n-2)-1$. Notice that $|S_2|=|\{(p,q,r)\in \mathbb{N}\times\mathbb{N}\times\mathbb{N}: 1\leq p \leq q\leq r, p+q+r=n-4\}|=P_3(n-4)$. Moreover, grounded on \cite{anderson_2006}, the number of partitions of $k$  with at most $3$ parts is equal to $\left[\frac{(k+3)^2}{12}\right]$.
It follows that $P_3(k)=\left[\frac{(k+3)^2}{12}\right]-\left\lfloor\frac{k}{2}\right\rfloor-1$ \cite{anderson_2006}.
Therefore, $$|S_2|=\left[\frac{(n-1)^2}{12}\right]-\left\lfloor\frac{n-4}{2}\right\rfloor-1$$ and $$|S-S_2|=\left\lfloor\frac{n-2}{2}\right\rfloor-1.$$
Thus, $$|S|=\displaystyle \left[\frac{(n-1)^2}{12}\right]-\left\lfloor\frac{n-4}{2}\right\rfloor+\left\lfloor\frac{n-2}{2}\right\rfloor-2.$$
{\hspace*{\fill}$\Box$\medskip}\\

\section{Conclusion}
The problems of finding the $(-k)$-critical graphs and the $k$-minimal graphs seem to be challenging where $k$ is an integer ($k\geq 2$). At least, we solve these problems in the particular case of trees. In addition, we determine the number of nonisomorphic $(-k)$-critical trees with $n\geq5$ vertices where $k \in \{1,2, \lfloor\frac{n}{2}\rfloor\}$. Besides, we specify the number of nonisomorphic $3$-minimal trees with $n$ ($n\geq 4$) vertices.

\nocite{*}
\bibliographystyle{abbrvnat}
\bibliography{sample-dmtcs}
\label{sec:biblio}

\end{document}